\documentclass[times, twoside]{zHenriquesLab-StyleBioRxiv}

\usepackage{amsmath, amsfonts, amssymb}  
\usepackage{lineno} 
\usepackage{listings} 
\usepackage{algorithm, algorithmic} 
\usepackage{graphicx} 
\usepackage{float} 
\usepackage{xcolor} 
\usepackage{cuted} 
\usepackage{array}
\usepackage{makecell}


\newcommand{\beginsupplement}{%
        \setcounter{table}{0}
       \renewcommand{\thetable}{S\arabic{table}}%
        \setcounter{figure}{0}
        \renewcommand{\thefigure}{S\arabic{figure}}%
     }

\leadauthor{Rich} 

\begin{document}


\title{Optimizing alluvial plots}
\author[1,2]{Joseph Rich}
\author[1]{Conrad Oakes}
\author[1,3]{Lior Pachter\textsuperscript{*}}

\affil[1]{Biology and Biological Engineering, California Institute of Technology, Pasadena, CA, 91125, USA}
\affil[2]{USC-Caltech MD/PhD Program, Keck School of Medicine, Los Angeles, CA, 90033, USA}
\affil[3]{Computing and Mathematical Sciences, California Institute of Technology, Pasadena, CA, 91125, USA}

\maketitle

\textsuperscript{*}Correspondence: lpachter@caltech.edu.


\section*{Abstract}
\noindent Alluvial plots can be effective for visualization of multivariate data, but rely on ordering of alluvia that can be non-trivial to arrange. We formulate two optimization problems that formalize the challenge of ordering and coloring partitions in alluvial plots. While solving these optimization problems is challenging in general, we show that the NeighborNet algorithm from phylogenetics can be adapted to provide excellent results in typical use cases. Our methods are implemented in a freely available R package available on GitHub at \url{https://github.com/pachterlab/wompwomp}

\textbf{Keywords:} alluvial plot, graph theory, greedy algorithm, NeighborNet


\section{Introduction}


\begin{figure*}[ht!]
    \centering
    \includegraphics[width=\textwidth]{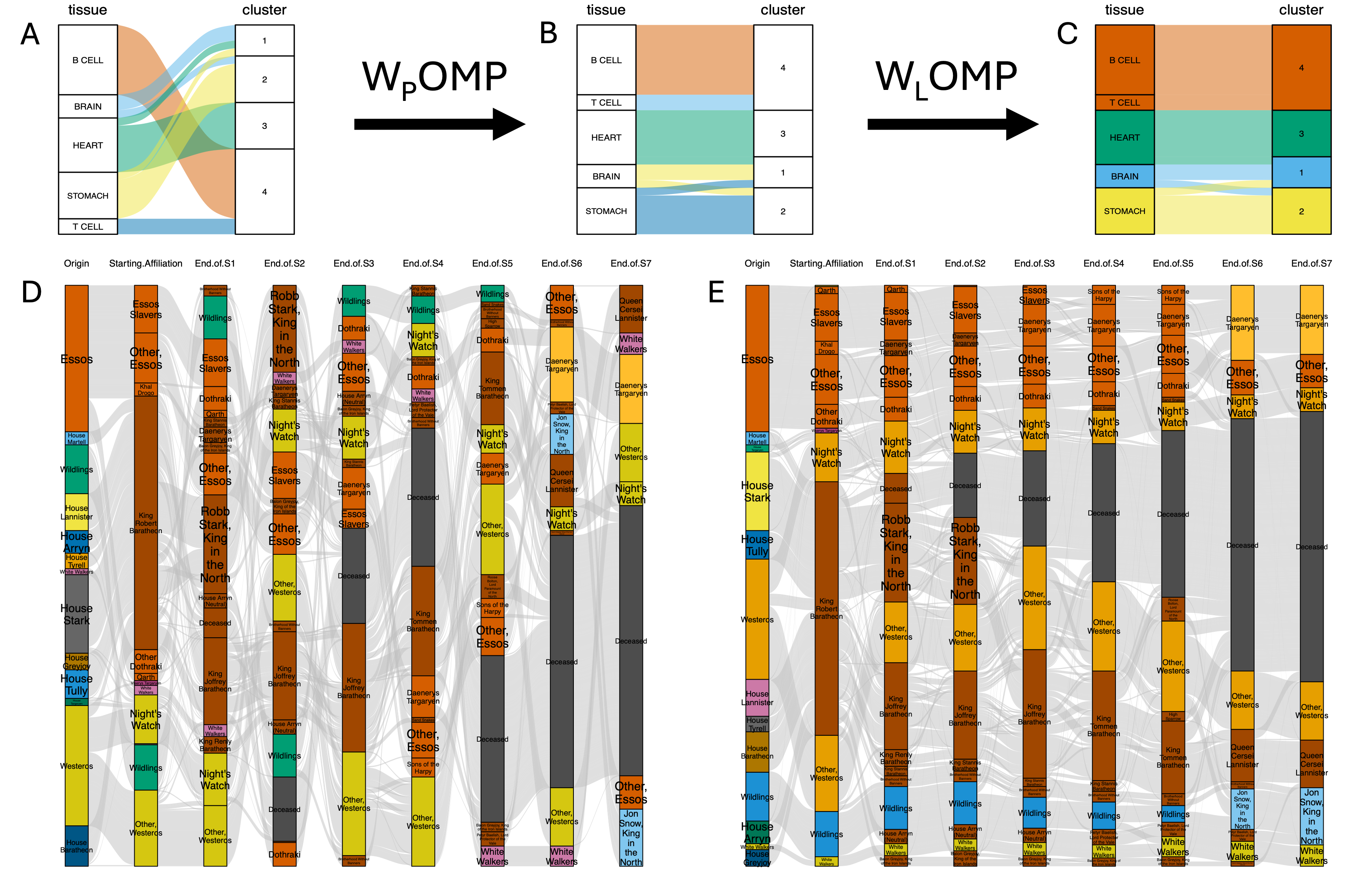}
    \caption{wompwomp overview. (A) Example with tissue to cluster mapping with randomized block partitions and coloring. (B) Example with tissue to cluster mapping with W\textsubscript{P}OMP applied (block sorting). (C) Example with tissue to cluster mapping with wompwomp applied (block sorting and color matching). (D) Example with Game of Thrones character political affiliation with randomized block partitions and coloring. (E) Example with Game of Thrones character political affiliation with wompwomp applied.}
    \label{fig:fig1}
\end{figure*}

A major challenge for exploratory analysis of multivariate data is the visualization of relationships among large numbers of variables. Alluvial plots or diagrams \cite{kennedy1898thermal,rosvall2010mapping} offer one approach to this challenge: categories are represented as strata within vertical bars, and individual observations are represented as lines or curves known as alluvia or flows. The alluvia connect strata across variables, displaying the category memberships of each observation for each variable (Fig. 1). 

Formally, an alluvial plot represents $n$ multivariate observations $X = \{x_1, x_2, \dots, x_n\}$ with respect to $m$ variables. The variables are represented by partitions $\Pi_1, \Pi_2, \dots, \Pi_m$ of $X$, where the blocks within each partition \( \Pi_i \) correspond to the categories for the respective variables:
\[
\Pi_i = \{B_1^{(i)}, B_2^{(i)}, \dots, B_{k_i}^{(i)}\}, \quad \text{with } \bigcup_{j=1}^{k_i} B_j^{(i)} = X.
\]

An alluvial plot provides a geometric realization of $X$ and the partitions $\Pi_1,\ldots,\Pi_m$ via specification of a permutation $\mu \in S_m$ defining the order of the variables, along with permutations $\sigma_i \in S_n^{(i)}$ ($i = 1, \dots, m$) specifying the order of the elements of $X$ for each variable. Crucially, alluvial plots require that the permutations $\sigma_i$ preserve the block structure of \( \Pi_i \), i.e., 
\begin{equation}
\label{eq:constraint}
\forall B_j^{(i)} \in \Pi_i, \max_{x \in B_j^{(i)}} \sigma_i(x) - \min_{x \in B_j^{(i)}} \sigma_i(x) + 1 = |B_j^{(i)}|.
\end{equation}
This ensures that the images of the elements of each block under each permutation must form a set of consecutive integers, so that that elements belonging to the same category of each variable are co-located in an alluvial plot.

To illustrate the structure of an alluvial plot and its associated definitions, consider a bivariate visualization task where $X$ consists of cells from a single-cell RNA-sequencing (scRNA-seq) experiment (Fig. 1A). To each observation we associate two variables: the first consists of categories that are tissue labels for the cells, and the second consists of categories that are cluster labels for cells, derived from a {\it de novo} clustering algorithm. In this example, $n=27$ and $m=2$ with $|\Pi_1| = 5$ and $|\Pi_2|=4$ (see also Supplementary Table 1). In an alluvial plot, the $m$ partitions are visualized as columns in the plot; these are also referred to as {\em layers}. Edges in the plot correspond to individual observations; these are also referred to as {\em alluvia}.

The utility of the alluvial representation is evident when comparing it to a single statistic that might be derived from the data. For example, the Adjusted Rand Index (ARI) can be used to measure the agreement between the two partitions $\Pi_1$ and $\Pi_2$ of the cells in the example and will consist of a single number ranging from -1 (complete incompatibility between partitions) to 1 (complete agreement between partitions) \citep{halkidiClusterValidityMethods2002}. However, the ARI is a summary that does not represent important information regarding the patterns of how individual cells are partitioned. Namely, ARI is not invariant to splitting or merging of elements between partitions. For instance, the fact that the block corresponding to ``4'' in $\Pi_2$ is largely composed of the ``T cell'' and ``B cell'' blocks from $\Pi_1$ is evident in the alluvial plot. Additionally, the ARI is sensitive to cluster splitting or merging, and therefore cannot capture information about superclusters or subclusters.

The observations in an alluvial plot frequently share metadata, and this can be formalized as follows: For each element \( x \in X \), we associate a weight \( w(x) \in \mathbb{R}_{>0} \), with the weight $w(x)$ corresponding to the multiplicity of the observation. The case where all weights are equal to one corresponds to the consideration of all observations individually. After collapsing, the number of elements $\overline{n}$ (visualized as unique paths in the plot) satisfies 
\[
\max_i (k_i) \leq \overline{n} \leq \min(n, K_{\text{prod}}),
\]
where $K_{\text{prod}} = \prod_{i=1}^{m} k_i$. Each layer $i$ has $|S_n^{(i)}| = \overline{n}!$ possible alluvium permutations (i.e., values for $\sigma_i$); across $m$ layers, there are $|S_n| = (\overline{n})^m$ possible alluvium permutations in total. However, this number includes invalid permutations of elements that do not preserve block structure. The number of possible valid permutations  \(|S_{n, valid}| = \prod_{i=1}^{m} \left( k_i! \cdot \prod_{j=1}^{k_i} \overline{n}_{i,j}! \right) \), where $\overline{n}_i,j$ is the number of alluvia in layer $i$ and block $j$. The problem of element partitioning can thus be reduced to block partitioning, where each layer $i$ has $|S_p^{(i)}| = k_i!$ possible block permutations, and an alluvial plot has in total $|S_p| = \prod_{i=1}^{m} k_i!$ possible block permutations. There are $|\mu| = m!$ variable permutations, and therefore $|S| = |\mu| \cdot |S_p|$ possible permutations to consider across both variables and blocks.


To elucidate the patterns in an alluvial plot, it is useful to consider two optimizations (Fig. 1A-C). First, optimization of the permutations $\mu$ and $\sigma_1,\ldots,\sigma_m$ can untangle the edges, thereby showcasing the overall concordance between variables (Fig. 1B). The permutation $\mu$ is, in current alluvial plotting tools, typically either determined by a temporal ordering of the variables or arbitrarily. The permutations $\sigma_1,\ldots,\sigma_n$, which specify the order of categories for each variable, are typically chosen manually or determined by the size of the blocks or alphabetical order \cite{ggalluvial}. Second, optimized color assignment to blocks in the partitions can help reveal which are most concordant (Fig. 1C). 

We formulate these optimization problems as follows:

For any two elements \( x, y \in X \), and adjacent partition indices \( r = 1, \dots, m-1 \), define
\[
\chi_{x,y}^{(r)} := 
\begin{cases}
1 & \text{if } 
\operatorname{sign}(\sigma_{\mu(r)}(x) - \sigma_{\mu(r)}(y)) \cdot {} \\[-0.5ex]
  & \quad\operatorname{sign}(\sigma_{\mu(r{+}1)}(x) - \sigma_{\mu(r{+}1)}(y)) = -1, \\
0 & \text{otherwise}.
\end{cases}
\]
Let 
\begin{equation}
\label{eq:loss}    
\mathcal{L}(\mu, \{\sigma_i\}) := \sum_{r=1}^{m-1} \sum_{\substack{x, y \in X \\ x \ne y}} w(x) \cdot w(y) \cdot \chi_{x,y}^{(r)}.
\end{equation}

The first optimization problem, which we refer to as W\textsubscript{P}OMP ({\bf W}eighted (permutation) {\bf O}ptimization (of) {\bf M}ultiple {\bf P}artitions), is to find $\mu \in S_m$ and permutations $\sigma_1, \dots, \sigma_m \in S_n$, each satisfying (\ref{eq:constraint}) that minimize $\mathcal{L}(\mu, \{\sigma_i\})$.

Next, let
\[
c^{(i)} : \{1, \dots, k_i\} \to \mathcal{C}
\]
be a bijection that assigns distinct colors (that is, non-negative integer labels) to each block in each partition. Define the weight of agreement between two blocks from adjacent partitions \( \Pi_i \) and \( \Pi_{i+1} \) as
\[
W_{j\ell}^{(i)} := \sum_{x \in B_j^{(i)} \cap B_\ell^{(i+1)}} w(x).
\]
Then, for each pair \( (\Pi_i, \Pi_{i+1}) \), define the total matched weight:
\[
\mathcal{M}^{(i)} := \sum_{\substack{j = 1,\dots,k_i \\ \ell = 1,\dots,k_{i+1}}} \delta\left(c^{(i)}(j), c^{(i+1)}(\ell)\right) \cdot W_{j\ell}^{(i)},
\]
where \( \delta(a,b) = 1 \) if \( a = b \), and \( 0 \) otherwise.
Finally, let 
\begin{equation}
\label{eq:2ndopt}
\mathcal{M} := \sum_{i=1}^{m-1} \mathcal{M}^{(i)}.
\end{equation}

The second optimization problem, which we refer to as W\textsubscript{L}OMP ({\bf W}eighted (label) {\bf O}ptimization (of) {\bf M}ultiple {\bf P}artitions) is to find assignments $\{c^{(i)}\}_{i=1}^m$ that maximize $\mathcal{M}$, i.e. the color agreement among variables.

The optimization of both the permutation determination problem and the label assignment problem is W\textsubscript{P}OMP-W\textsubscript{L}OMP. The W\textsubscript{P}OMP problem is a  generalization of the Weighted One-Layer Free Problem (WOLF). The WOLF task is to minimize the products of overlapping edges in a bipartite graph with the ordering of one layer fixed \citep{cakirogluCrossingMinimizationWeighted2009}. W\textsubscript{P}OMP extends WOLF in that it applies to a multipartite graph where no layers are fixed. The WOLF problem is NP-hard, and therefore the W\textsubscript{P}OMP is NP-hard.

While W\textsubscript{P}OMP is NP-hard, we developed an approach that improves the loss function (\ref{eq:loss}) for W\textsubscript{P}OMP in the small tissue-cluster mapping example from 182 to 1 (Fig. 1A-B). Such improvements are not always achieved because $S_p$ can grow rapidly when the number of blocks and / or the number of variables increases. An example of a more complex application of alluvial plotting is in tracking shifting political affiliations of characters in Game of Thrones across time \citep{lunkes_game_2019}. In this example, $n = 488$, $m = 9$, $\overline{n} = 183$ leading to $K_{sum} = 122$, and $K_{prod} = 12,546,293,760$ where $K_{\text{sum}} = \sum_{i=1}^{m} k_i$ (the total number of blocks across all variables) and 
$K_{\text{prod}} = \prod_{i=1}^{m} k_i$ (the total number of possible block combinations across all variables). In this example, where the order of variables represents a meaningful quantity (time), it does not make sense to optimize $\mu$, but even without that optimization the problem of minimizing edge crossing is challenging. In this case, our approach improves the loss function in the Game of Thrones example from 343,087 to 62,905 (Fig. 1D-E).

We address W\textsubscript{P}OMP by applying the Neighbornet algorithm \citep{bryantNeighbornetAgglomerativeMethod2004, levyNeighbornetAlgorithm2011} leading to an algorithm that runs in $\left(K_{\text{sum}}^3 + K_{\text{sum}} \cdot m^2 \cdot \overline{n}\log \overline{n}\right)$ time and $O\left(K_{\text{sum}}^2 + \overline{n} + m^2 \right)$ space, 
with optimizations put in place to improve both time and space performance in the vast majority of cases. The W\textsubscript{L}OMP problem is implemented by performing hierarchical clustering of blocks based on element overlap, and runs in \(O(m \cdot K_{\text{sum}}+K_{\text{sum}} \cdot log(K_{\text{sum}}))\) time and \(O(K_{\text{sum}}^2)\) space. These functions are implemented in the R package wompwomp (Supplementary Fig. 1).

\section{W\textsubscript{P}OMP}
The first step in wompwomp that addresses W\textsubscript{P}OMP is to run the NeighborNet algorithm \cite{bryantNeighbornetAgglomerativeMethod2004} on a graph derived from $X$ and its associated $m$ partitions (Algorithm 1). The intuition motivating the use of NeighborNet is that it is a greedy algorithm for finding circular split systems of minimal balanced length \cite{levyNeighbornetAlgorithm2011}, which in the alluvial plot application translates to finding a cycle between layers that reduces crossovers (Fig. 2, Supplementary Fig. 2).

\begin{figure*}[ht!]
    \centering
    \includegraphics[width=\textwidth]{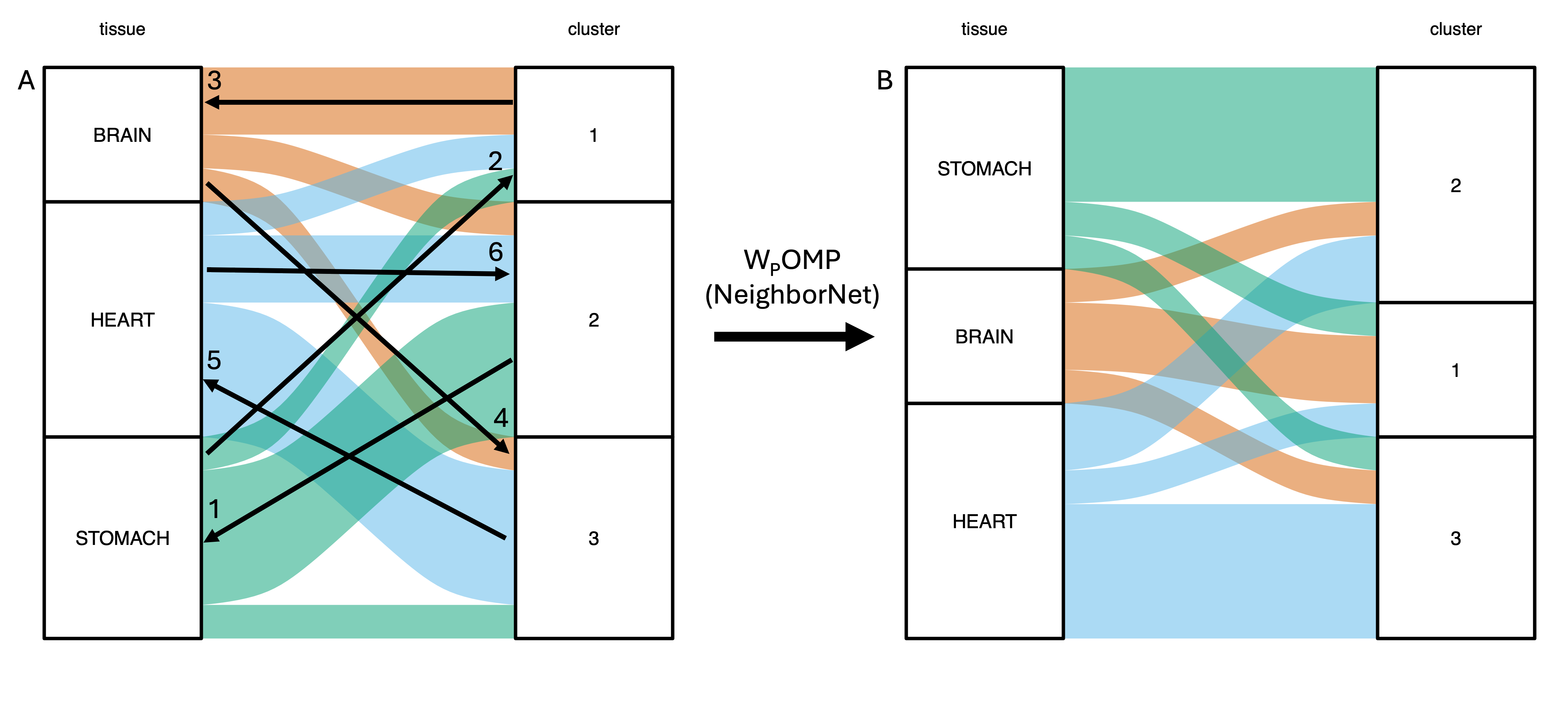}
    \caption{Walkthrough of wompwomp on a two-layer example. (A) Initial alluvial plot with randomized block partitions and no block coloring. Numbered arrows indicate the cycle produced by W\textsubscript{P}OMP (NeighborNet), where the path follows the arrow from tail to head; numbers correspond to the optimal starting position. (B) Alluvial plot after applying W\textsubscript{P}OMP (NeighborNet) sorting.}
    \label{fig:fig2}
\end{figure*}

The weighted graph we construct can be represented as a symmetric \( K_{\text{sum}} \times K_{\text{sum}} \) distance matrix, where each entry \( (i, j) \) is the value \( c \cdot \left( -\log(w_{ij}) \right) \), with \( w_{ij} \) denoting the weight between blocks \( i \) and \( j \) and $c$ denoting a positive scalar multiplier which is a parameter. This scalar multiplier $c$ acts to modulate the effect of differences in edge weight when creating the cycle. Blocks are ordered based on a flattened list of all \( K_{\text{sum}} \) blocks across layers. This matrix takes $O(K_{\text{sum}}^2)$ space. We initialize this matrix with large values. Optionally, we can initialize values for nodes within the same layer as a different value than for nodes in different layers. We then create a data frame where each row represents a combination of variables and the subsequent weight of this combination, which takes $O(\overline{n})$ space. We loop through this data frame, filling the matrix with the minus log of the edge weights. This entire process takes \( O\left(m^2 \cdot \overline{n} \log \overline{n}\right) \) time. We then run NeighborNet on this distance matrix as implemented in SplitsPy (\url{https://github.com/husonlab/SplitsPy}). NeighborNet runs in \( O(K_{\text{sum}}^3)\) time. The combined matrix construction and NeighorNet algorithm runs in \( O\left(m^2 \cdot \overline{n} \log \overline{n} + K_{\text{sum}}^3\right) \) time. For small $m$ relative to $K_{\text{sum}}$, which is the case in most applications where $m$ is between 2 and 10, this simplifies to \( O(K_{\text{sum}}^3)\) time. The output of the NeighborNet function is a cycle that lists all $K_{\text{sum}}$ nodes in sorted order.

Although the resultant cycle lists the nodes in sorted order relative to each other, this does not necessarily indicate where the optimal starting point of the cycle is when flattened into a multipartite graph. In order to determine a starting point, we loop through each potential starting point $i$ of the $K_{\text{sum}}$ options, split this cycle into its $m$ layers, optionally determine the optimal order of layers with determine\_layer\_order (Algorithm 2), and then determine the objective function $\mathcal{L}(\mu_i, \{\sigma_i\})$ with compute\_objective (Algorithm 3). Our final output is the graph $(\mu_i, \{\sigma_i\})$ with the smallest objective function. We have found that, generally, the optimal order of layers does not change with different cycle starting points, and therefore we only perform this step in the first iteration of the loop. The time complexity of this algorithm is \( O\bigl(\text{determine\_layer\_order}(m,\overline{n}) 
   + K_{\text{sum}} \cdot m \cdot \text{compute\_objective}(\overline{n}) \bigr) \).

In order to calculate the objective function $\mathcal{L}(\mu_i, \{\sigma_i\})$ with compute\_objective (Algorithm 3), we utilize a Fenwick tree (binary indexed tree) approach. For all $m-1$ pairs of adjacent layers, we loop through all $\overline{n}$ rows in the input dataframe, first sorting them by their position in the source layer ($y_1$), and then using a rank-compressed version of the target layer ($y_2$) as the indexing coordinate in the tree. As we iterate over the sorted rows, we use the tree to maintain a prefix sum of edge weights and efficiently query the cumulative weight of edges that would cross the current edge (i.e., those with higher $y_2$). This allows us to incrementally compute the total crossing weight in $O(\overline{n} \log \overline{n})$ time per layer pair, with $O(\overline{n})$ space, enabling scalable optimization of $\mathcal{L}$ even for large input graphs. We sum the objective for each pair of adjacent layers, yielding an overall time complexity of $O(m \cdot \overline{n} \log \overline{n})$.

In order to determine the optimal layer order with determine\_layer\_order (Algorithm 2), we run a traveling salesman problem (TSP) solving algorithm on an \( m \times m \) distance matrix. The motivation of using a TSP solver is to order layers such that similar layers end up near each other, which reduces to finding an optimal Hamiltonian cycle where each node represents a variable, and each edge represents a distance value. Each entry of the distance matrix stores \(c \cdot \log\left(1 + \mathcal{L}(\mu_i, \{\sigma_i, \sigma_j\})\right) \), where $c$ denotes a positive scalar multiplier. This objective function is computed even if $i$ and $j$ are non-adjacent. The $log1p$ function is applied rather than $log$ to avoid $log(0)$ errors. We calculate this objective function on each of the \( \binom{m}{2} \) combinations of layers, where each objective function is computed in $O(\overline{n}\log \overline{n})$ time complexity as described above. The TSP solver of arbitrary insertion with two-opt refinement is run on the resulting matrix, which runs in $O(m^2)$ time (two-opt refinement is a constant time operation for symmetric matrices). We then determine the starting point in the cycle by placing the largest adjacent distance at the end of the tour to minimize visual discontinuity. This results in an overall time complexity of \( O\left(m^2 \cdot \overline{n}\log \overline{n} \right) \) and space complexity of \( O\left(m^2 + \overline{n}\right) \). An alternative metric to store in this \( m \times m \) matrix is the ARI, which can be calculated in $O(n + K_{sum}^2)$ for each pair of layers, yielding overall time complexity of \( O\left(m^2 \cdot (n + K_{sum}^2) \right) \) and space complexity of \( O\left(m^2 + n\right) \).

Plugging in these results, this yields an overall time complexity for determining the optimal cycle starting point of \( O\left( (m^2 \overline{n}\log \overline{n}) + K_{\text{sum}} \cdot (m^2 \cdot \overline{n}\log \overline{n}) \right) \), or simply \( O\left( K_{\text{sum}} \cdot (m^2 \overline{n}\log \overline{n}) \right) \). Combining this with the NeighborNet algorithm described previously, this yields \( O\left(K_{\text{sum}}^3 + K_{\text{sum}} \cdot m^2 \cdot \overline{n}\log \overline{n}\right) \) time complexity and $O\left(K_{\text{sum}}^2 + \overline{n} + m^2\right)$ space complexity for W\textsubscript{P}OMP overall.

\section{W\textsubscript{L}OMP}
The core algorithm to maximize (\ref{eq:2ndopt}) for W\textsubscript{L}OMP operates on two layers, with one used as the potential parent. For each block of the child layer, its overlap with each block in the parent layer is calculated as
\[
B_p^{(i)}\cap B_c^{(j)}, 
\]
where \(B_p^{(i)}\) is the $i$th block of the parent layer and \(B_c^{(j)}\) is the $j$th block of the child layer. In this way, each potential parent block gets a score between 0 and 1. 

From here, two potential schemes are offered to resolve color matching. In the first, every potential pair of parent and child layers are calculated. The resulting matrix can be thought of as a graph where each node is a block and the connection weight is the parent score between that block and another. This graph is then clustered via Leiden or Louvain methods to determine similar blocks \citep{traag_louvain_2019}. Each cluster of similar blocks is then assigned the same color. The calculation of parent scores has a time complexity of \(O(m \cdot K_{\text{sum}})\), while the Leiden clustering has a time complexity of \(O(K_{\text{sum}} \cdot log(K_{\text{sum}}))\). This leaves the overall time complexity as \(O(m \cdot K_{\text{sum}}+K_{\text{sum}} \cdot log(K_{\text{sum}}))\). The largest data structure during this algorithm is a data frame with at most one row for each pair of blocks in the graph, resulting in \(O(K_{\text{sum}}^2)\) space complexity.

The second option is to first set a layer as a reference layer, then for each layer, to assign colors based off which block in the reference layer has the highest parent score (optionally also requiring a minimum parent value). The reference layer can either be held constant, or can vary throughout the alluvial plot starting from either the left or right. 

\section{Applications}
\subsection{Alluvial visualization of evolving political affiliations in Game of Thrones}

Various sorting schemes exist to permute blocks for representation in alluvial plots including random order ($O(n)$), alphabetical order ($O(n\log n)$), descending size ($O(n\log n)$), and wompwomp (described earlier) (Supplementary Fig. 3). In the Game of Thrones example, random sorting yields an objective of 343,087, descending order 193,466, alphabetical order 123,924, and wompwomp 62,905. Descending and alphabetical order perform significantly better than random order because there is shared information between layers. Because characters generally remain in constant affiliations between seasons, this means that the blocks, with names that remain similar among layers, will have contents that correlate with the name. Additionally, if characters generally stay where they are, then corresponding blocks will stay at a roughly similar size across layers and tend to line up. However, both of these heuristics substantially underperform in comparison to wompwomp. 

The utility of wompwomp can be demonstrated by highlighting trajectories of specific affiliations throughout the show, such as those of Lannister and Westeros. In the case of random permuting, these affiliations cross over each other frequently, and it is difficult to tell if this crossover is a result of frequent swapping of affiliations or an artifact of suboptimal sorting (Supplementary Fig. 4A). After applying W\textsubscript{P}OMP, we can more clearly visualize that Lannister and Westeros do indeed have generally separate and consistent trajectories. A better alluvial plot also makes it easier to track the fraction of characters from each starting affiliation who join the deceased group by the end of each season (Supplementary Fig. 4B). Moreover, it is easier to tell which starting affiliations end up in Lannister or Westeros by the end of the show when the plot is generally sorted to straighten out these trajectories (Supplementary Fig. 4C-D).

\subsection{Differential gene expression in scRNA-seq}
To illustrate the application of wompwomp to genomics data, we applied it to a scRNA-seq dataset in which gene expression was measured across 8 tissues in 8 mouse strains, with 8 biological replicates for each combination \citep{rebboah_systematic_nodate}. When plotting the alluvia for two genes, the objective function improved substantially with wompwomp: from 60,652,796,431 billion in the unsorted view to 28,071,624,548 billion after sorting alluvia (Fig. 3A-C). The resulting visualization reveals interpretable biological structure. For example, the Slc7a12 gene shows higher expression in female mice, while the Akr1c21 gene is biased towards male mice. Slc7a12 is also more prevalent in strains such as PWKJ and CASTJ, and less so in B6 and NODJ. The persistent crossover between the sex and genotype layers indicates that both genes are expressed in all strains, highlighting a level of entanglement that cannot be resolved by sorting alone.

\begin{figure*}[ht!]
    \centering
    \includegraphics[width=\textwidth]{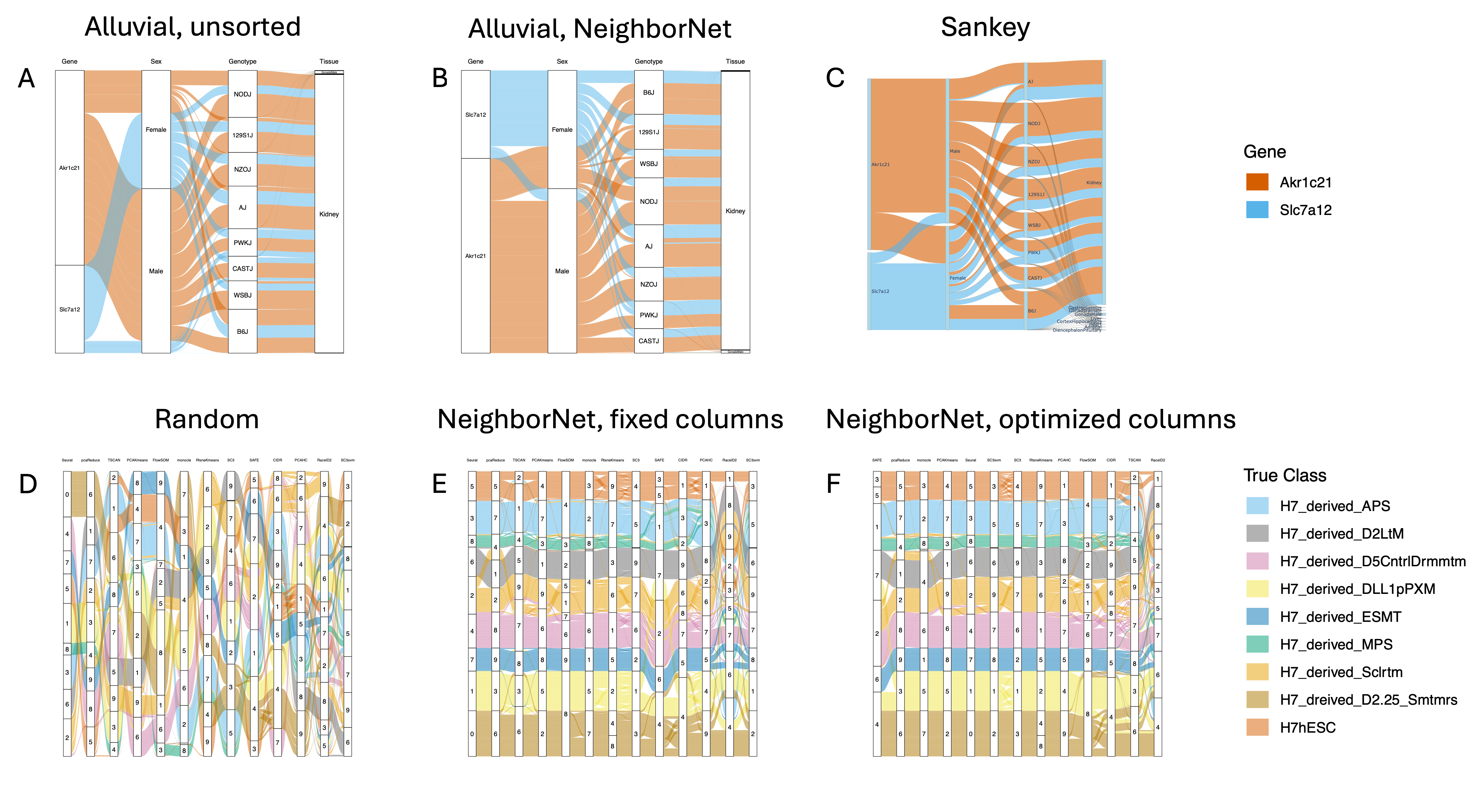}
    \caption{wompwomp example on scRNA-seq data. (A) Alluvial plot of two genes with unsorted partitions and unmatched colors. (B) Alluvial plot with sorted partitions and matched colors using wompwomp. (C) Sankey diagram. (D) Clustering dataset alluvial plot with unsorted partitions and unmatched colors, fixed layers. (E) Clustering dataset alluvial plot with sorted partitions and matched colors using wompwomp, fixed layers (F) Clustering dataset alluvial plot with sorted partitions and matched colors using wompwomp, optimized layers. Top legend shows color mapping for panels A-C. Bottom legend shows color mapping for panels D-F.}  
    \label{fig:fig3}
\end{figure*}

\subsection{Comparison of clustering results}
In order to demonstrate a use case where layer order optimization is relevant, we visualized the 13 unsupervised clustering methods applied to a dataset of 531 cells with predetermined ground truth labels \citep{duo_systematic_2020}. With random block permutations, the alluvial plot had an objective of 625,967 (Fig. 3D). With wompwomp and fixed layer order, the objective improved to 101,256 (Fig. 3E). With wompwomp and optimized layer order, the objective improved further to 56,374 (Fig. 3F). With optimized layer order we can observe that the layers near the ends of the alluvial plot such as RaceID2\citep{grun_novo_2016}, TSCAN\citep{ji_tscan_2016}, CIDR\citep{lin_cidr_2017}, and SAFE\citep{yang_safe-clustering_2019} generally have the most disagreements with other methods; whereas the methods near the middle of the alluvial plot such as RtsenKmeans\citep{van2014accelerating}, SC3\citep{kiselev_sc3_2017}, SC3svm,\citep{cortes_support-vector_1995} and Seurat\citep{satija_spatial_2015} generally have the most agreement (Fig. 3F).

Overall, these results are in agreement with Figure 4 of \citep{duo_systematic_2020}, which hierarchically clustered the methods and visualized their similarities via a tree. Specifically, RaceID2 is identified as an outlier in both approaches, although some of the details in terms of which methods are most similar differ a bit. Importantly though, the hierarchical clustering of \citep{duo_systematic_2020} does not reveal how truth assignments of each cell are predicted by each method, something that the alluvial plot in Figure 3F does well. 

\subsection{Alluvial plot-based comparison of machine learning model performance and demographic effects}
To visualize and compare the performance of the two machine learning models on the KiTS19 test set \citep{raman_evaluation_2025}, we applied our optimized alluvial sorting algorithm to the Dice scores generated by each model across the 90 test cases. 

Supplementary Figure 5A displays the unsorted flow between Dice score bins from the KiTS-trained model (Model 1) and the USC-trained model (Model 2), while Supplementary Figure 5B shows the same data with the alluvial sorting algorithm applied. The sorting reduced the objective function from 467 to 407. The sorted version reveals clear trends: Model 1 consistently outperforms Model 2. Notably, approximately 10\% of test images achieve Dice scores in the 0.9–1.0 range for Model 1 but fall in the 0.0–0.1 range for Model 2, and an additional 2\% of images show 0.8–0.9 Dice for Model 1 but again drop to 0.0–0.1 for Model 2. The large block in the bottom-left corner of the flow diagram (Model 2 Dice 0.0–0.1) with no corresponding block in Model 1 further highlights the underperformance of the USC model.

We next stratified Dice score performance by demographic and clinical factors to examine potential sources of variation. Supplementary Figure 5C–D shows Model 1’s performance across gender, ethnicity, and tumor size. The sorting reduced the objective function from 1288 to 1081. There was no visible difference in performance across ethnicity or tumor size, but a modest drop in performance for female patients compared to male patients was observed. In contrast, Supplementary Figure 5E–F shows that Model 2's performance was significantly affected by tumor size, with the model struggling more on small tumors than large ones. No notable differences were seen across ethnicity or gender. The sorting for Supplementary Figure 5E-F reduced the objective function from 2156 to 1762.

\begin{algorithm}
\caption{\textsc{W\textsubscript{P}OMP}\_NeighorNet}
\begin{algorithmic}[1]
\REQUIRE A dataframe \texttt{df} with $m+1$ columns: $m$ categorical variables (\texttt{graphing\_columns}) and one column specifying the count for each unique combination of variable values.

\STATE Let $\mathcal{K} \gets$ the ordered list of all $K_{\text{sum}}$ unique values across \texttt{graphing\_columns}
\STATE Initialize a $K_{\text{sum}} \times K_{\text{sum}}$ matrix $M$ with large values \COMMENT{$M[i, j]$ corresponds to distance between $\mathcal{K}[i]$ and $\mathcal{K}[j]$}
\STATE Construct a long-format dataframe \texttt{df\_long} with columns: \texttt{variable1}, \texttt{variable2}, \texttt{value}
\FOR{each row in \texttt{df\_long}}
    \STATE Let $i_{name} \gets$ value of \texttt{variable1} in row
    \STATE Let $j_{name} \gets$ value of \texttt{variable2} in row
    \STATE Let $v \gets$ value of \texttt{value} in row
    \STATE Set $M[i_{name}, j_{name}] \gets c \cdot -\log(v)$ \COMMENT{$c > 0$ is a positive scalar multiplier}
\ENDFOR
\STATE Let $cycle \gets$ NeighborNet$(M)$ \COMMENT{Circular ordering of the $K_{\text{sum}}$ values}
\STATE \texttt{graphing\_columns\_optimized} $\gets$ \texttt{graphing\_columns}
\STATE \texttt{df\_best} $\gets$ \texttt{df}
\STATE \texttt{objective\_best} $\gets \infty$
\STATE \texttt{graphing\_columns\_optimized\_best} $\gets$ \texttt{graphing\_columns\_optimized}

\FOR{each possible starting point in $cycle$}
    \STATE Split $cycle$ into $m$ contiguous sublists, one per variable/layer
    \FOR{each graphing column $k = 1$ to $m$}
        \STATE \texttt{df[\textquotesingle col\_k\_int\textquotesingle]} $\gets$ rank of \texttt{df[\textquotesingle col\_k\textquotesingle]} based on order in sublist $k$
    \ENDFOR
    \IF{layer order optimization is enabled}
        \STATE \texttt{graphing\_columns\_optimized} $\gets$ \texttt{determine\_layer\_order(df, graphing\_columns)}
    \ENDIF
    \STATE \texttt{objective} $\gets$ \texttt{compute\_objective(df, graphing\_columns\_optimized)}
    \IF{\texttt{objective} $<$ \texttt{objective\_best}}
        \STATE \texttt{df\_best} $\gets$ \texttt{df}
        \STATE \texttt{objective\_best} $\gets$ \texttt{objective}
        \STATE \texttt{graphing\_columns\_optimized\_best} $\gets$ \texttt{graphing\_columns\_optimized}
    \ENDIF
\ENDFOR

\RETURN a tuple:
\begin{itemize}
    \item \texttt{df\_best} — the dataframe with new columns \texttt{col\_k\_int} for $k = 1, \ldots, m$, where each \texttt{col\_k\_int} maps values in \texttt{graphing\_columns[$k$]} to their position in the corresponding layer
    \item \texttt{graphing\_columns\_optimized\_best} — the final optimized layer order
\end{itemize}
\end{algorithmic}
\end{algorithm}

\begin{algorithm}[H]
\caption{determine\_layer\_order}
\begin{algorithmic}[2]
\REQUIRE Dataframe \texttt{df} with $m$ columns (\texttt{col1\_int}, ..., \texttt{colm\_int}) and \texttt{value}
\STATE Initialize an $m \times m$ matrix $D$ with zeros
\FOR{each pair of layers $(i, j)$ with $1 \leq i < j \leq m$}
    \STATE Compute $\mathcal{L}(\mu_i, \{\sigma_i, \sigma_j\})$ \COMMENT{Objective function between layers $i$ and $j$, in $O(\overline{n}\log \overline{n})$ time}
    \STATE $D[i, j] \gets D[j, i] \gets c \cdot \log\left(1 + \mathcal{L}(\mu_i, \{\sigma_i, \sigma_j\})\right)$ \COMMENT{$c > 0$ is a positive scalar multiplier; use $\log1p$ to avoid $\log(0)$}
\ENDFOR
\STATE Let $\pi \gets$ TSP($D$) \COMMENT{Run TSP on $D$ to obtain optimal ordering $\pi$ over $m$ layers}
\STATE $\pi \gets \operatorname{rotate\_left}(\pi, \arg\max_i D[\pi_i, \pi_{(i \bmod n) + 1}])$  \COMMENT{Rotate $\pi$ left to place the largest adjacent distance at the end}
\RETURN $\pi$
\end{algorithmic}
\end{algorithm}

\begin{algorithm}[H]
\caption{compute\_objective}
\begin{algorithmic}[3]
\REQUIRE Dataframe \texttt{df} with columns \texttt{col1\_int}, $\ldots$, \texttt{colm\_int}, and \texttt{value}
\STATE \texttt{df\_alluvia} $\gets$ \texttt{get\_alluvia\_df(df)} \COMMENT{Contains $\overline{n}$ alluvia; performed by ggalluvial}
\STATE \texttt{objective} $\gets$ 0
\FOR{each adjacent layer $k = 1$ to $m-1$}
    \STATE \texttt{df\_pair} $\gets$ \texttt{df\_alluvia[[col$k$\_int, col$k$+1\_int]]} renamed as \texttt{y1}, \texttt{y2}
    \COMMENT{df\_pair contains columns y1, y2, count}
    \STATE Sort \texttt{df\_pair} by \texttt{y1} ascending
    \STATE Compute dense ranks \texttt{y2\_rank} of \texttt{y2} (larger $\rightarrow$ higher rank)
    \STATE Initialize BIT of size equal to $\max(\texttt{y2\_rank}) + 2$
    \FOR{each row $i$ in \texttt{df\_pair}}
        \STATE \texttt{weight} $\gets$ \texttt{count[i]}, \texttt{rank} $\gets$ \texttt{y2\_rank[i]}
        \STATE \texttt{sum\_above} $\gets$ BIT.query\_range(\texttt{rank + 1}, \texttt{max\_rank})
        \STATE \texttt{objective} $\gets$ \texttt{objective} $+$ \texttt{weight} $\times$ \texttt{sum\_above}
        \STATE BIT.update(\texttt{rank}, \texttt{weight})
    \ENDFOR
\ENDFOR
\RETURN \texttt{objective}
\end{algorithmic}
\end{algorithm}

\begin{algorithm}[H]
\caption{\textsc{W\textsubscript{L}OMP}}
\begin{algorithmic}[4]
\REQUIRE Dataframe \texttt{df} with columns \texttt{col1\_int}, $\ldots$, \texttt{colm\_int}, and \texttt{value}
\FOR{each pair of layers $(i, j)$}
    \FOR{each pair of blocks $(B_i, B_j)$}
        \STATE Compute $B_i \cap B_j $ \COMMENT{Overlap between block in layer $i$ and block in layer $j$}
    \ENDFOR
\ENDFOR
\end{algorithmic}
\end{algorithm}



\section{Discussion}

This work was motivated by previous work in which we considered the problem of how to best render alluvial plots to compare unsupervised clustering output \citep{richImpactPackageSelection2024b}. We initially implemented a greedy algorithm that kept one layer fixed, reordered blocks in the other layer so as to match the thickest edges to the opposite side, and colored blocks using maximum bipartite matching. However, our greedy WOLF algorithm, with $O(k_1 \cdot k_2)$ time complexity (where $k_i$ is the number of blocks in layer $i$), had some major drawbacks. The method was not suitable for multivariate problems with more than two variables, and even in the case of two variables it was not suitable for applications where there is no intrinsic ordering of either layer. To address these caveats, we employed the NeighborNet algorithm, originally designed for applications in phylogenetics as an alternative to Neighbor Joining (Supplementary Fig. 6). The intuition guiding the use of NeighborNet was that the algorithm can be viewed as a sophisticated greedy algorithm with an optimality guarantee for Kalmanson metrics, which could be common in alluvial plot applications where minimizing crossovers is a desired goal. The coaxing of the alluvial plot representation problem into a NeighborNet application involved construction of a distance matrix representing the inverse weight between blocks, i.e., \(W[i,j] = c \cdot \left( -\log(w_{ij}) \right) \). The parameter $c$ tunes the extent to which NeighborNet favors large weights over small weights when determining optimal cycles.

This algorithm and visualization approach has broad utility across domains. We have demonstrated its effectiveness in tracking unique molecular identifier (UMI) distributions in scRNA-seq, visualizing character affiliations across seasons in a television show, and comparing machine learning model performance across demographic groups or models. Beyond these applications, it is well-suited for comparing clustering outputs by revealing how groupings from one method map onto another and highlighting areas of agreement or disagreement. It also applies naturally to energy flow diagrams, where conservation and distribution of energy types can be traced across systems \cite{schmidt_sankey_2008}. More generally, we believe that wompwomp will be valuable for general multivariate categorical data analyses, including patient treatment trajectories \cite{ottoOverviewSankeyFlow2022}, task allocations in workflow pipelines, storytelling \cite{shixia_liu_storyflow_2013, tanahashiDesignConsiderationsOptimizing2012}, and evolutionary transitions in phylogenetic classifications. Additionally, the x- or y-axes can optionally carry physical meaning: for example, using time on the x-axis enables tracking changes over time, while assigning anatomical position to the y-axis (e.g., tissue depth) allows blocks to reflect spatial organization, such as placing outer layers at the top and deeper tissues below.

The different nature of Sankey diagrams also has implications for how the loss function would be computed. The loss function between any two layers $(i, i+1)$ in both alluvial plots and Sankey diagrams is computed in $\overline{n}_i \log \overline{n}_i$ operations, where $\overline{n}_i$ is the number of alluvia between the two layers and $1 \leq i \leq m - 1$. However, rather than in alluvial plots where $\overline{n}_i = \overline{n} \quad \forall\, i$ and $\overline{n} \leq \min(n, K_{\text{prod}})$, for Sankey diagrams, it is instead $\overline{n}_i \leq k_i \cdot k_{i+1}$. The number of alluvia is not fixed across all layers, but rather  changes as we observe different pairs of adjacent layers. The objective function for both alluvial plots and Sankey diagrams can be calculated in $\sum_{i=1}^{m-1} \overline{n}_i \log \overline{n}_i$ operations. For alluvial plots, this simplifies to $(m-1) \cdot \overline{n} \log \overline{n}$, which is $O(m \cdot \overline{n} \log \overline{n})$. However, this does not reduce for Sankey diagrams, leading to $O(\sum_{i=1}^{m-1} \overline{n}_i \log \overline{n}_i) = O(m \cdot \hat{\overline{n}} \log \hat{\overline{n}} \quad \text{where } \hat{\overline{n}} = \max_i \overline{n}_i)$. $\hat{\overline{n}} \leq \overline{n}$, as $\hat{\overline{n}}$ has an upper bound of the product of blocks in any two adjacent layers, whereas $\overline{n}$ has an upper bound of the product of blocks across all layers $K_{prod}$. Note that $\hat{\overline{n}} = \overline{n}$ if and only if the alluvial plot has at most two layers with more than one block --- more formally, $\hat{\overline{n}} = \overline{n} \iff \left| \left\{ i \in \{1, \dots, m\} \mid k_i > 1 \right\} \right| \leq 2$.

While alluvial diagrams and Sankey plots are often used interchangeably to visualize flows between categories, they differ in both structure and interpretability. An example of each is shown in Figure 3B-C. One of the distinctions lies in the treatment of the vertical (y-axis) position. In alluvial diagrams, the y-axis has quantitative meaning: the height of each block or stratum corresponds to a cumulative count or value. The position of each alluvium is directly determined by its order and size, reflecting the distribution of elements in the dataset. In contrast, Sankey plots allow arbitrary spacing between flows and strata. These visual gaps are often introduced for aesthetic clarity, but they decouple the diagram’s layout from the actual data distribution. Another important difference lies in how flow paths are represented. Alluvial plots track each alluvium as a distinct path across all layers, allowing us to follow a unique combination of variables from start to end. This preserves the identity of each element or group of elements, which is especially valuable when each unit is meaningful—for instance, tracking individual patients or images in a radiology dataset. By contrast, Sankey plots only represent the aggregated flow between adjacent layers. Individual paths are not preserved across the entire plot, which reduces visual clutter but loses information about how data partitions across multiple stages. The tradeoff is between clarity and completeness. Sankey plots typically exhibit fewer crossovers, producing a cleaner and more visually appealing layout. However, the reduced complexity comes at the cost of interpretability: Sankey plots cannot show how a single element flows across more than two dimensions, nor do they reveal how combinations of variables persist or split over time. Alluvial diagrams, while more complex and potentially messier due to crossing paths, provide a richer picture of the dataset’s structure. They allow us to understand not just the distribution of values in each layer, but how those distributions interact, align, or diverge across dimensions.

Our method has several notable limitations. First, it can struggle with one-to-many node relationships: since the inferred cycle visits each node only once, only two edges (one in, one out) are considered, leaving additional connections unaccounted for. While this could theoretically degrade performance, it appears to have little effect in practice. Second, manual inspection sometimes reveals small changes that improve the layout. A promising strategy may be to use the current approach (e.g., NeighborNet) as an initialization, followed by a greedy refinement step that guarantees not to worsen the objective. Third, because the algorithm operates solely on a distance matrix between blocks, it lacks access to any initial structure or user-supplied priors, which limits control over starting configurations. Finally, the cyclic nature of the output makes it difficult to fix a particular layer in place; doing so would break assumptions of the algorithm. 

\section{Conclusion}
We defined the W\textsubscript{P}OMP-W\textsubscript{L}OMP problem as a partitioning problem of elements under a constraint of maintaining block continuity, followed by subsequent coloring to match similar groups. We developed a heuristic algorithm for solving W\textsubscript{P}OMP by applying the NeighborNet algorithm, as well as for solving W\textsubscript{L}OMP by clustering blocks by similar parent overlap scores. These algorithms and their visualization with alluvial plots are implemented in the open source R package wompwomp.



\section*{Acknowledgments}
We thank Nikhila Swarna for help with testing examples for Fig. 3A,B,C. Thanks to Ruth Liorsdóttir for suggesting the name wompwomp for the software package.

\section*{Author Contributions}
J.R., C.O., and L.P. conceived the project. J.R. wrote the W\textsubscript{P}OMP NeighborNet algorithm and additional preprocessing code. C.O. wrote the W\textsubscript{L}OMP algorithm and plotting code. J.R. and C.O. wrote the software and vignettes. J.R. wrote the initial version of the manuscript. J.R., C.O., and L.P. read and edited the manuscript.

\section*{Declaration of Interests}
None declared.


\section*{Methods}
wompwomp is an open-source R package ($R \geq 3.5.0$) that implements a heuristic solution for the W\textsubscript{P}OMP-W\textsubscript{L}OMP problem and plots the results with the ggalluvial package \citep{ggalluvial}. It is available on GitHub (\url{https://github.com/pachterlab/wompwomp}) and has been submitted to Bioconductor. It supports both an R interface and a command line interface, where any function FUNCTION in R can be run on the command line with /Path/to/wompwomp/exec/wompwomp FUNCTION. It is designed to interoperate with standard data.frame structures. Core data manipulation is performed using the dplyr, purrr, and tidyr packages, while visualization builds upon ggplot2 using extensions including ggalluvial, ggforce, and ggfittext \citep{tidyverse, ggalluvial}.

\subsection*{Data preprocessing}
The core data input to wompwomp consists of either:

an $n \times m$ data frame, where each row represents an entity and each column a categorical variable; or

an $\overline{n} \times (m + 1)$ data frame, where each row represents a unique combination of values across the $m$ grouping variables, along with an additional column encoding the number of entities sharing that combination.

The function \texttt{wompwomp::data\_preprocess()} supports both input forms. It standardizes the input by:
\begin{itemize}
    \item mapping string categories to integer identifiers according to the value of ``default\_sorting'' (for sorting and plotting),
    \item inserting a default \texttt{``Missing''} label for \texttt{NA} values, and
    \item transforming raw per-row data (first form) into a grouped data frame (second form).
\end{itemize}

default\_sorting can take on the following values:
\begin{itemize}
    \item 'alphabetical' (default)
    \item 'reverse\_alphabetical'
    \item 'increasing'
    \item 'decreasing'
    \item 'random'
\end{itemize}

The resulting grouped data frame includes integer-encoded columns for each categorical variable and is used downstream by the sorting and plotting functions.

\subsection*{Sorting Algorithms}

Ordering of strata within each layer (W\textsubscript{P}OMP) is managed by \texttt{wompwomp::data\_sort()}, which takes the grouped form of the data and returns an updated data frame with (potentially re-mapped) integer layers reflecting the new order. Four options are available via the \texttt{method} argument:

\begin{itemize}
    \item \texttt{``none''}: No sorting is applied; strata retain their original order. If an original order is not present, then it is added via data\_preprocess and the default\_sorting argument.
    \item \texttt{``random''}: Randomized assignment.
    \item \texttt{``greedy\_WOLF''}: A greedy heuristic for the Weighted One Layer Free (WOLF) bipartite ordering problem, as described in Rich et al. \citep{richImpactPackageSelection2024b}. One layer is fixed; the other is reordered to minimize edge crossings by assigning each group to align with the group in the fixed layer that shares the thickest edge.
    \item \texttt{``greedy\_WBLF''}: A symmetric extension of WOLF, where both layers are iteratively optimized, treating both layers as free. This also applies only to bipartite graphs.
    \item \texttt{``neighbornet''}: A more general ordering using the W\textsubscript{P}OMP algorithm described previously. This method utilizes the NeighborNet algorithm from the \texttt{SplitsPy} Python package, which constructs a circular ordering of the variables that reflects pairwise distances. A key customization in \texttt{wompwomp} initializes the pairwise distance matrix with zeros (rather than uninitialized memory via \texttt{np.empty}), improving stability.
    \item \texttt{``tsp''}: Works like ``neighbornet'', but constructs the cycle via TSP implemented from the \texttt{TSP} R package \citep{hahsler_tspinfrastructure_2008} rather than via NeighborNet. 
\end{itemize}

The \texttt{neighbornet} algorithm is accessed from R via the \texttt{reticulate} package. The Python environment can be set up via \texttt{wompwomp::setup\_conda\_env()} as either a conda environment (default) or virtualenv.

\subsection*{Visualization}

The wompwomp method is run using the wrapper function \texttt{wompwomp::plot\_alluvial()} which run the individual wompwomp steps by calling \texttt{data\_preprocess()}, \texttt{data\_sort()}, and \texttt{plot\_alluvial\_internal()}. The internal plotting function constructs the alluvial plot by interpreting the sorted and grouped data frame as a graph and rendering it using \texttt{ggalluvial} \citep{ggalluvial}. Color matching (W\textsubscript{L}OMP) is performed within \texttt{plot\_alluvial\_internal}.

\newpage
\section*{References}
\bibliography{ref}

\clearpage
\onecolumn        
\section*{Supplementary Information}
\beginsupplement
\begin{figure*}[ht!]
    \centering
    \includegraphics[width=\textwidth]{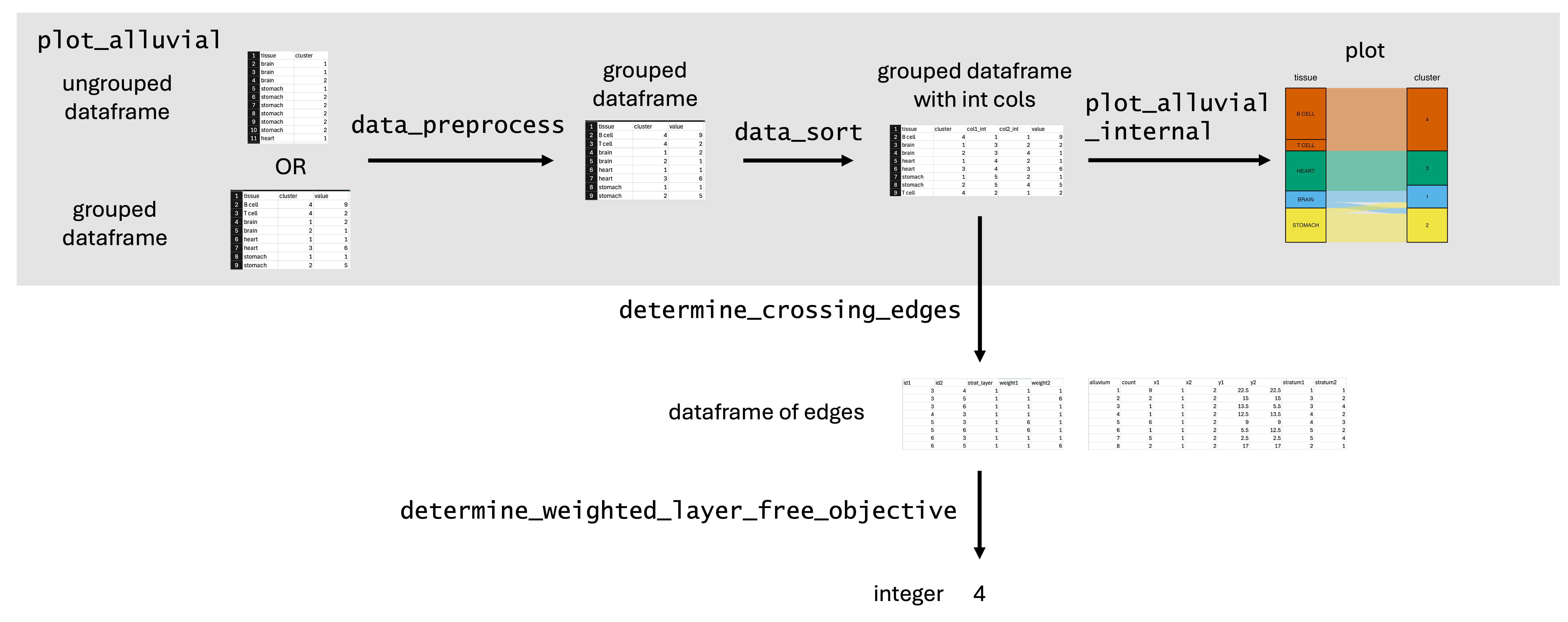}
    \caption{wompwomp schematic. Input is a grouped or ungrouped dataframe. data\_preprocess performs preprocessing on a grouped or ungrouped datagrame. data\_sort applies the W\textsubscript{P}OMP step. plot\_alluvial\_internal applies the W\textsubscript{L}OMP step and produces the resulting plot. plot\_alluvial wraps data\_preprocess, data\_sort, and plot\_alluvial\_internal. determine\_crossing\_edges determines crossing edges in the alluvial plot, and determine\_weighted\_layer\_free\_objective calculates the sum products of overlapping edge weights.}  
    \label{fig:supp1}
\end{figure*}
\newpage

\begin{figure*}[ht!]
    \centering
    \includegraphics[width=\textwidth]{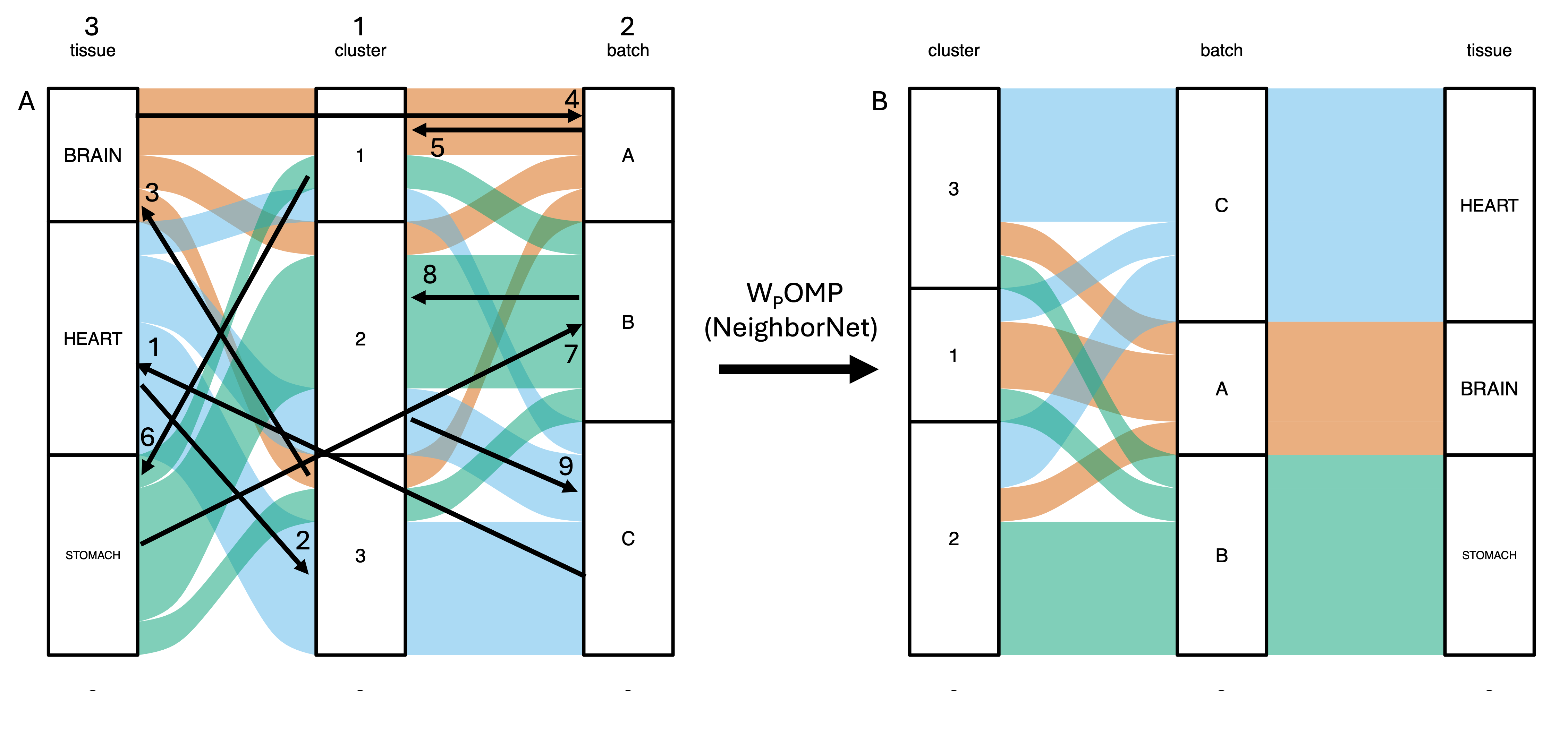}
    \caption{Walkthrough of wompwomp on a three-layer example. (A) Initial alluvial plot and induced graph with randomized block partitions and no block coloring. Numbered arrows indicate the cycle produced by W\textsubscript{P}OMP (NeighborNet), where the path follows the arrow from tail to head; numbers above arrows correspond to the optimal starting position of the cycle; numbers above layers correspond to the optimal layer ordering. (B) Alluvial plot after applying W\textsubscript{P}OMP (NeighborNet) sorting.}
    \label{fig:supp2}
\end{figure*}
\newpage

\begin{figure*}[ht!]
    \centering
    \includegraphics[width=\textwidth]{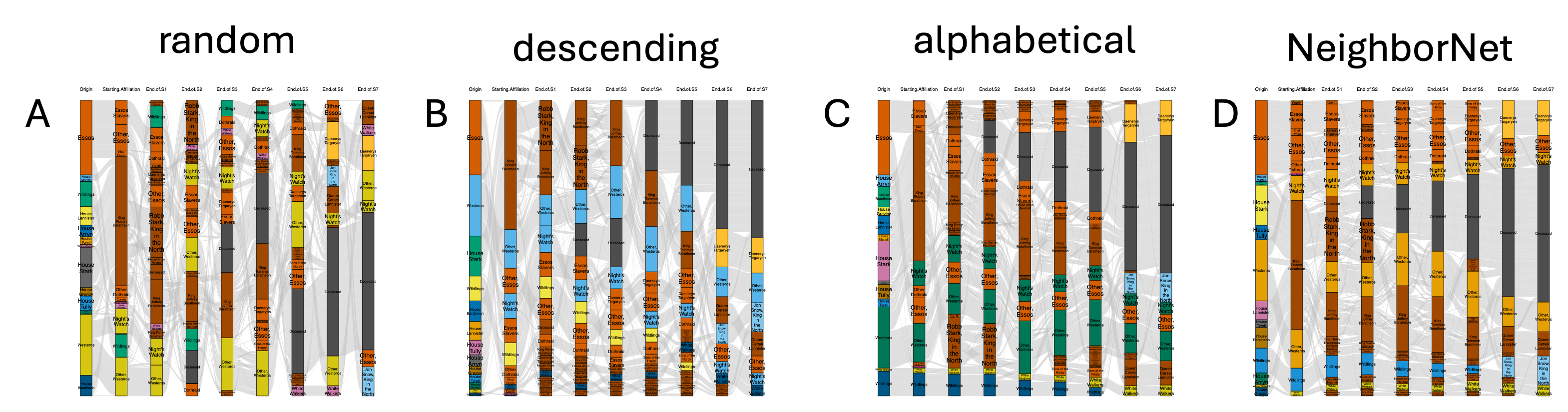}
    \caption{Game of Thrones affiliations example with additional alternative partitioning algorithms. (A) Random order (same as Fig. 1D). (B) Descending order by block size. (C) Alphabetical order. (D) wompwomp (same as Fig. 1E).}  
    \label{fig:supp3}
\end{figure*}
\newpage

\begin{figure*}[ht!]
    \centering
    \includegraphics[width=\textwidth]{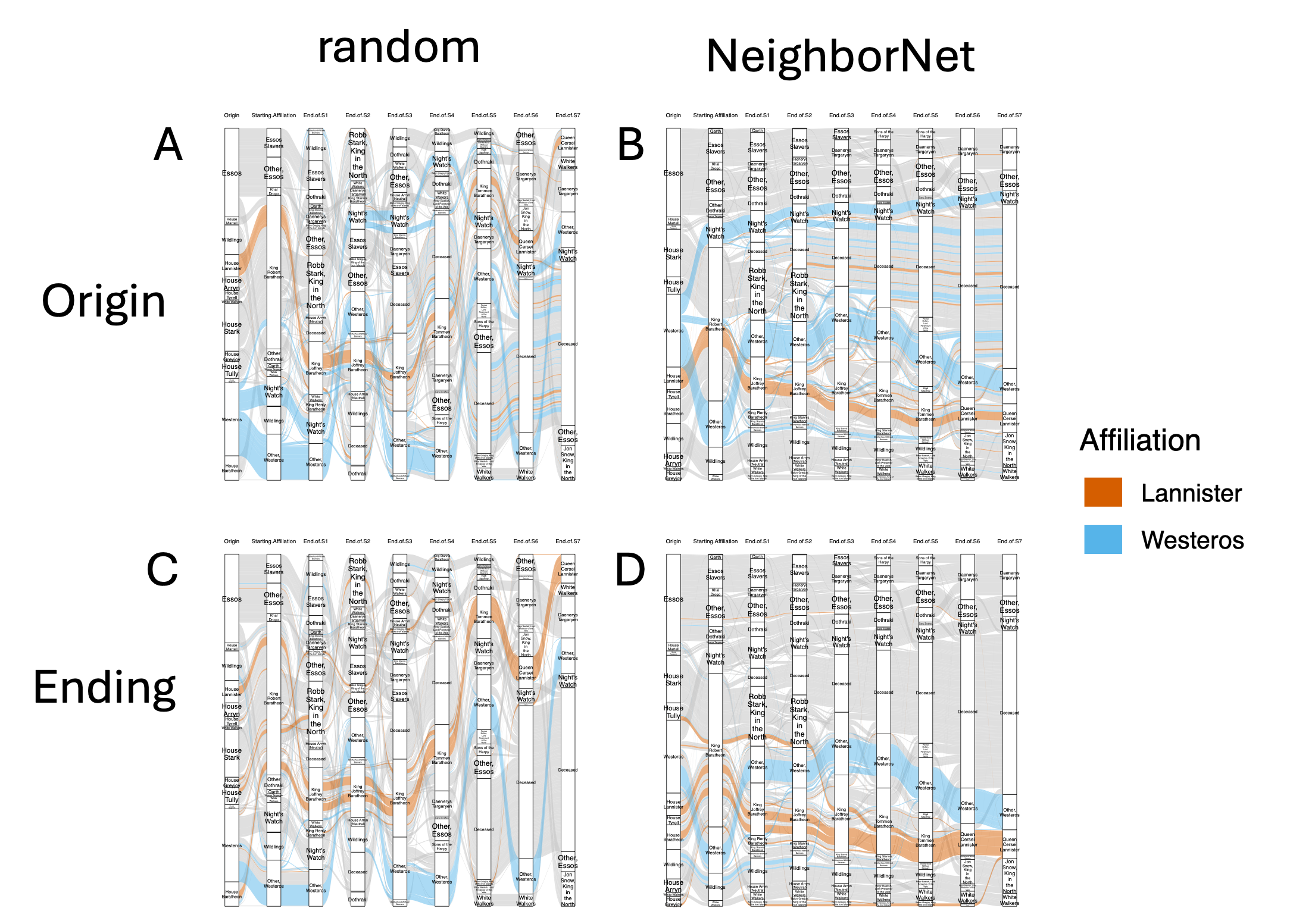}
    \caption{Game of Thrones affiliations example, band coloring by affiliation. (A) Unsorted partitions, colored by affiliation at origin. (B) Sorted partitions with wompwomp, colored by affiliation at origin. (C) Unsorted partitions, colored by affiliation at ending. (D) Sorted partitions with wompwomp, colored by affiliation at ending. Orange = Lannister, blue = Westeros.}  
    \label{fig:supp4}
\end{figure*}
\newpage

\begin{figure*}[ht!]
    \centering
    \includegraphics[width=\textwidth]{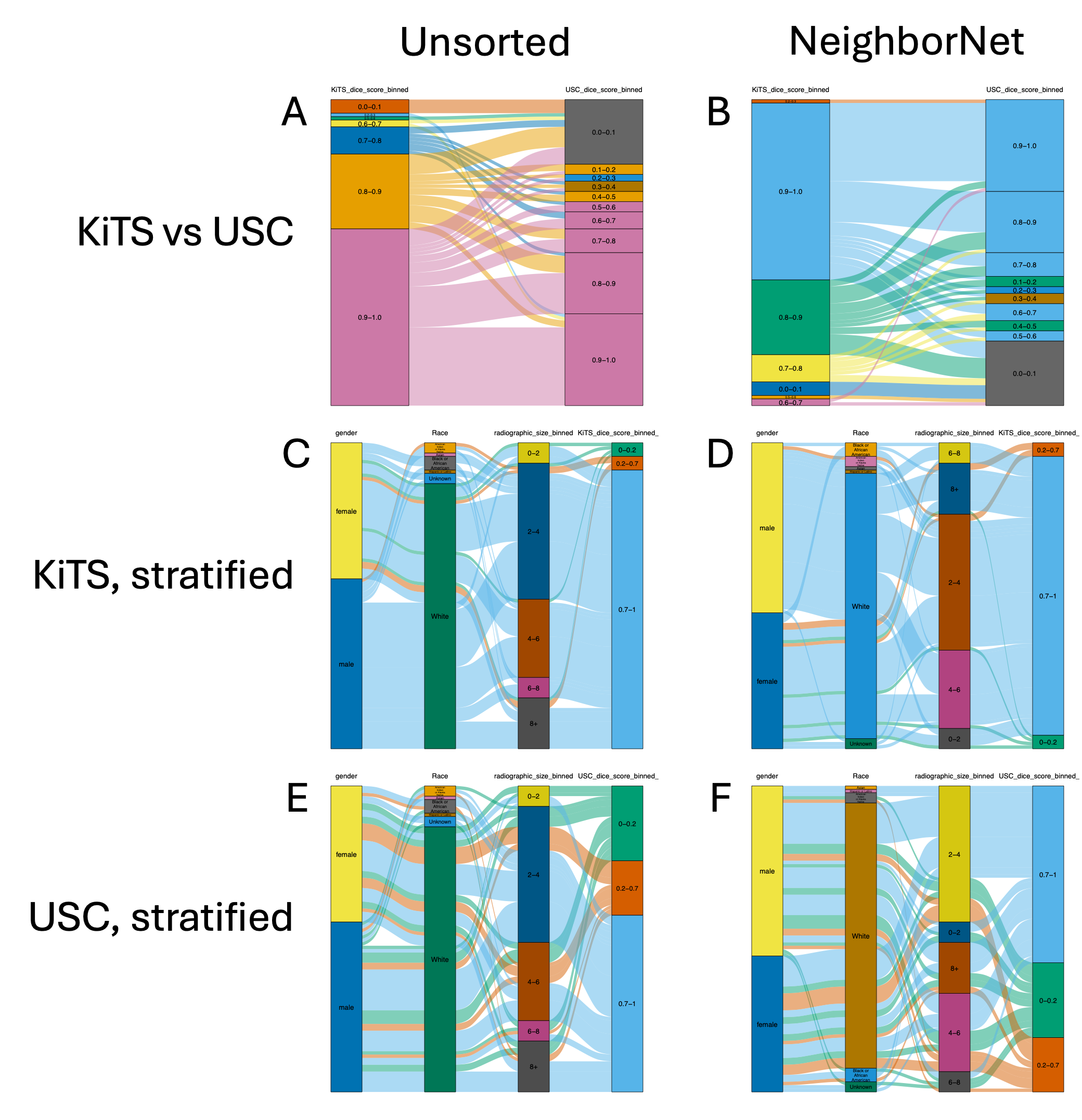}
    \caption{Machine learning model comparison. (A) Model1 vs. Model2, unsorted. (B) Model1 vs. Model2, wompwomp. (C) Model1 stratified by demographic factors, unsorted. (D) Model1 stratified by demographic factors, wompwomp. (E) Model2 stratified by demographic factors, unsorted. (F) Model2 stratified by demographic factors, wompwomp. In C-F, coloring = Dice score range.}  
    \label{fig:supp5}
\end{figure*}
\newpage

\begin{figure*}[ht!]
    \centering
    \includegraphics[width=\textwidth]{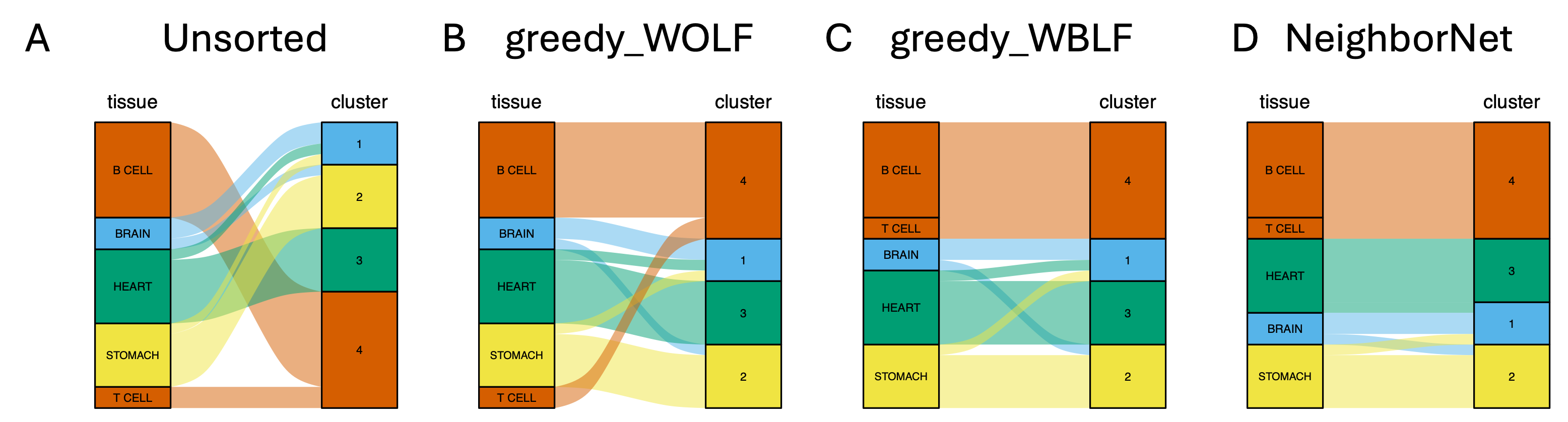}
    \caption{Tissue-cluster mapping example, different sorting algorithms. (A) Unsorted. (B) greedy\_WOLF. (C) greedy\_WBLF. (D) wompwomp (with NeighborNet).}  
    \label{fig:supp6}
\end{figure*}
\newpage


\newpage
\begin{table}[ht]
\centering
\begin{tabular}{|l|p{2.4cm}|p{4.7cm}|p{5.1cm}|}
\hline
\textbf{Symbol} & \textbf{Value} & \textbf{Equation} & \textbf{Description} \\
\hline
$n$ & 27 & N/A & Number of elements \\
$m$ & 2  & N/A & Number of layers   \\
$\overline{n}$ & 8  & N/A & Number of alluvia  \\
$k_1$  &  5  &  N/A   & Number of blocks in layer 1 \\
$S_p^{(1)}$  &  120  &  $k_1!$   &  Number of block permutations in layer 1 \\
$\Pi_1$  & \makecell[l]{\{B CELL,\\ BRAIN,\\ HEART,\\ STOMACH,\\ T CELL\}} & $\{B_1^{(1)}, B_2^{(1)}, B_3^{(1)}, B_4^{(1)}, B_5^{(1)}\}$ & Block permutation in layer 1 \\
$k_2$  &  4  &  N/A   &  Number of blocks in layer 2 \\
$S_p^{(2)}$  &  24  &   $k_2!$  &  Number of block permutations in layer 2  \\
$\Pi_2$  & \{1, 2, 3, 4\} & $\{B_1^{(2)}, B_2^{(2)}, B_3^{(2)}, B_4^{(2)}\}$ & Block permutation in layer 2 \\
$K_{sum}$  &  9  &   $\sum_{i=1}^{m} k_i$  &  Number of blocks across all layers \\
$K_{prod}$  &  20  &  $\prod_{i=1}^{m} k_i$   &  Number of combinations of blocks across layers  \\
$|\mu|$  &  2  &   $m!$  &   Number of permutations of layers   \\
$|S_p|$  &  2880  &  $\prod_{i=1}^{m} k_i!$   &  Number of permutations of blocks  \\
$|S|$  &  5760  &   $|\mu| \cdot |S_p|$  &    Number of permutations of layers and blocks  \\      
\hline
\end{tabular}
\caption{Model parameters and notation applied to Fig 1A-C.}
\label{tab:model_params}
\end{table}
\newpage

\end{document}